\documentclass[aps,prc,preprint,amsmath,amssymb,showpacs,preprintnumbers,superscriptaddress]{revtex4-1}
\usepackage{CJK}
\usepackage{graphicx}% Include figure files
\usepackage{dcolumn}% Align table columns on decimal point
\usepackage{bm}% bold math
\usepackage{color}
\usepackage{hyperref}% add hypertext capabilities
\allowdisplaybreaks[4]

\begin{document}

\title{Toward a bridge between relativistic and nonrelativistic density functional theories for nuclei}% Force line breaks with \\
\author{Z. X. Ren}
\affiliation{State Key Laboratory of Nuclear Physics and Technology, School of Physics, Peking University, Beijing 100871, China}

\author{P. W. Zhao}
\email{pwzhao@pku.edu.cn}
\affiliation{State Key Laboratory of Nuclear Physics and Technology, School of Physics, Peking University, Beijing 100871, China}

\begin{abstract}
The nonrelativistic reduction of the self-consistent covariant density functional theory is realized for the first time with the similarity renormalization group (SRG) method.
The reduced nonrelativistic Hamiltonian and densities are calculated by solving the corresponding flow equations with a novel expansion in terms of the inverse of the Dirac effective mass.
The efficiency and accuracy of this newly proposed framework have been demonstrated for several typical spherical nuclei.
It is found that the exact solutions of the total energies, traces of vector and scalar densities, and the root-mean-square radii are reproduced quite well for all nuclei.
This allows one to directly compare and bridge the relativistic and nonrelativistic nuclear energy density functional theories in the future.
\end{abstract}

%\pacs{21.60.Jz, 21.10.-k, 21.10.Re, 27.20.+n}
% 21.60.Jz Nuclear Density Functional Theory and extensions
%21.10.-k Properties of nuclei; nuclear energy levels
%21.10.Re Collective levels
%21.60.Ev Collective models
%21.60.Cs   Shell model
%23.20.-g Electromagnetic transitions
%23.20.Js Multipole matrix element
%27.20.+n  6 A 19
%27.60.+j 90  A 149
%27.50.+e 59  A  89

\maketitle

%*********************************************************%
%---------------------Introduction------------------------%
%*********************************************************%
%\section{Introduction}
The experimental facilities with radioactive beams have extended our knowledge of nuclear chart to the very limits of nuclear binding.
Theoretically, enormous efforts have been made to understand the physics of nuclear many-body systems based on microscopic approaches.
The nuclear density functional theory (DFT) is one of the most popular approaches in this context~\cite{Bender2003Self}. It starts from
a universal energy density functional and can achieve a global description for almost all nuclei in a fairly good accuracy.

Depending on the way how the energy density functionals are built, there are two kinds of DFTs for nuclear systems including the nonrelativistic and covariant DFTs.
Due to the consideration of the Lorentz symmetry, the covariant density functional theory (CDFT) has attracted a lot of attention in nuclear physics~\cite{meng2016relativistic}.
An essential ingredient of CDFT is to solve the relativistic Kohn-Sham equation with its effective single-particle potentials, i.e., a Dirac equation with an attractive scalar potential and a repulsive vector potential.
Both potentials are as large as several hundreds MeV, while the cancellation between them leads to a relatively weak potential felt by nucleons in the Fermi sea.
This reveals clearly the relativistic dynamics, rather than the relativistic kinematics, in describing the phenomena of low-energy nuclear structure. For this reason, the nonrelativistic density functionals are also very successful.

To bridge the nonrelativistic and covariant DFTs has been a longstanding task in nuclear physics, and it would allow a direct comparison between the nonrelativistic limit of  covariant density functionals and the nonrelativistic density functionals.
Such a comparison is motivated by the fact that the role and importance of the various terms in either covariant or nonrelativistic energy density functionals have not been completely understood so far. Furthermore, different density functional predictions exhibit systematic differences, which cannot yet be mapped onto the corresponding features of energy density functionals. A bridge between the nonrelativistic and covariant DFTs could help clarify these questions.

In particular, the CDFT allows one to describe the spin-orbit coupling and the time-odd fields~\cite{Vretenar2005PhysicsReport, Meng2013FT_TAC} in a natural and consistent way.
This could provide useful guidance for nonrelativistic density functionals.
For instance, the spin-orbit properties in CDFT have been used to improve the isospin dependence of the spin-orbit potential in nonrelativistic Skyrme functionals~\cite{Reinhard1995Pb_shift}.
Such a connection is also helpful to deepen our understanding on the interplay between the scalar and vector channels in the CDFT.

An efficient nonrelativistic reduction of the Dirac equation with large scalar and vector potentials is crucial to bridge the covariant and nonrelativistic DFTs.
This is usually achieved by solving the Sch\"{o}dinger-like equations for the upper and lower components of the Dirac spinors~\cite{Reinhard1989RPP,meng1998pseudospin}.
Based on this framework, it was found that the density dependence of the standard Skyrme functional may be too simple, and the expansion of the folding in terms of $\Delta\rho$ may be inappropriate~\cite{Reinhard1989RPP}.

In Ref.~\cite{bylev1998SRG}, a novel procedure for continuous unitary transformations, known as the similarity renormalization group (SRG) method, was introduced to reduce the Dirac Hamiltonian to a quasidiagonal form, i.e., two noninteracting parts corresponding to the positive and negative energies, respectively.
The flow equations of the reduced nonrelativistic Hamiltonian are solved by an expansion in a series of $1/M$ ($M$ is the bare mass of the Dirac particle).
In contrast to the Sch\"{o}dinger-like equations, the Hermitian of the reduced nonrelativistic Hamiltonian here is guaranteed at every order of the expansion.
This method has been used to investigate the pseudospin symmetries in nuclei~\cite{Guo2012SRG_spherical, Guo2014SRG_def} and the nuclear proton radioactivity~\cite{Zhao2014ProtonRadioactivity}.
However, due to the existence of the large scalar potential in CDFT, the convergence of the $1/M$ expansion is very slow.
Specifically, to acquire an accuracy around 0.2 MeV for the single-particle energies, the expansion should be at least up to $1/M^4$ order~\cite{Guo2019SRG}.
This prevents it from self-consistent calculations.

In the very recent work, an efficient solution of the Dirac Hamiltonian flow equations has been proposed through a novel expansion with the inverse of the Dirac effective mass $\tilde{M}=M+S$~\cite{Ren2019SRG}.
It is found that the exact solutions of the Dirac equation can be reproduced with a high accuracy up to only a few lowest order terms in the expansion.
This provides the possibility to perform self-consistent CDFT calculations with the SRG method.
However, to achieve the goal, one has to build the scalar and vector densities accurately with the wavefunctions at the nonrelativistic limit by solving the corresponding flow equations.

In this paper, for the first time, the covariant DFT is solved self-consistently by a nonrelativistic reduction with the SRG method.
The flow equations for the Dirac Hamiltonian and the vector- and scalar-density operators are solved with a novel expansion in terms of the inverse of the Dirac effective mass $1/\tilde{M}$.
The efficiency and accuracy of the nonrelativistic reduced CDFT are demonstrated for typical spherical nuclei.

%==================Theoretical framework===================%
%\section{Theoretical framework}
In CDFT, one needs to solve the Dirac Hamiltonian,
\begin{equation}\label{Eq_Dirac_hamiltonian}
  H=\bm{\alpha}\cdot\bm{p}+\beta(M+S)+V,
\end{equation}
where $\bm{\alpha}$ and $\beta$ are the Dirac matrices, $M$ is the mass of nucleon, and $S$ and $V$ are the scalar and vector potentials, respectively.
With the SRG method~\cite{bylev1998SRG, wegner1994SRG}, the Dirac Hamiltonian $H$ can be transformed by a unitary operator $U(l)$ as
\begin{equation}\label{Eq_hamiltonian_l}
   H(l)=U(l)HU^\dag(l),~~H(0)=H,
\end{equation}
where $l$ is a flow parameter.
The flow equation can be obtained by calculating the derivative of $H(l)$ with respect to $l$,
\begin{equation}\label{Eq_flow_eq}
  \frac{dH(l)}{dl}=[\eta(l),H(l)],
\end{equation}
with the generator,
\begin{equation}
   \eta(l)=\frac{dU(l)}{dl}U^\dag(l)=-\eta^\dag(l).
\end{equation}
Similar to Ref.~\cite{Ren2019SRG}, the generator $\eta(l)=[\beta,H(l)]$ is chosen to transform the Dirac Hamiltonian [Eq.~\eqref{Eq_Dirac_hamiltonian}] into a block-diagonal form.

The Hamiltonian $H(l)$ can be written as an even operator $\mathcal{E}(l)$ and an odd one $\mathcal{O}(l)$,
\begin{equation}\label{Eq_H_division}
  H(l)=\mathcal{E}(l)+\mathcal{O}(l),
\end{equation}
where the even and odd operators are defined by the commutation relation with the $\beta$ matrix, i.e., $\mathcal{E}(l)\beta=\beta\mathcal{E}(l)$ and $\mathcal{O}(l)\beta=-\beta\mathcal{O}(l)$.
The initial conditions for $\mathcal{E}(l)$ and $\mathcal{O}(l)$ read,
\begin{equation}\label{Eq_inital_O}
   \mathcal{E}(0)=\beta(M+S)+V,~~\mathcal{O}(0)=\bm{\alpha}\cdot\bm{p}.
\end{equation}
In Ref.~\cite{Ren2019SRG}, both $\mathcal{E}(l)$ and $\mathcal{O}(l)$ are solved by a perturbative expansion of $1/\tilde{M}$ with $\tilde{M}=M+S$.
One can find that $\mathcal{E}(l)$ is finite and $\mathcal{O}(l)$ exponentially goes to zero in the limit of $l\rightarrow\infty$.
Therefore, one can finally obtain a  block-diagonal Hamiltonian  $H(\infty)$.

In the following, we denote the eigenfunctions of the original Hamiltonian $H$ in Eq.~\eqref{Eq_Dirac_hamiltonian} and the transformed one $H(\infty)$ as $\psi$ and $\phi$, respectively.
Obviously, they are connected by the unitary transformation $U(l)$,
\begin{equation}
  \psi=\lim_{l\rightarrow\infty} U^\dag(l)\phi.
\end{equation}
Therefore, the vector and scalar densities are written as,
\begin{subequations}
  \begin{align}
    \rho_v(\bm{r})&=\int d^3r'~\psi^\dag(\bm{r}')\delta(\bm{r}-\bm{r}')\psi(\bm{r}')\nonumber\\
    &=\lim_{l\rightarrow\infty}\int d^3r'~\phi^\dag(\bm{r}')U(l)\delta(\bm{r}-\bm{r}')U^\dag(l)\phi(\bm{r}'),\label{Eq_vector}\\
    \rho_s(\bm{r})&=\int d^3r'~\psi^\dag(\bm{r}')\beta\delta(\bm{r}-\bm{r}')\psi(\bm{r}')\nonumber\\
    &=\lim_{l\rightarrow\infty}\int d^3r'~\phi^\dag(\bm{r}')U(l)\beta\delta(\bm{r}-\bm{r}')U^\dag(l)\phi(\bm{r}').\label{Eq_scalar}
  \end{align}
\end{subequations}
By defining the transformations,
\begin{equation}
  R_v(l)\equiv U(l)\delta(\bm{r}-\bm{r}')U^\dag(l),~~R_s(l)\equiv U(l)\beta\delta(\bm{r}-\bm{r}')U^\dag(l),
\end{equation}
one can obtain the flow equations of $R_{v,s}(l)$,
\begin{equation}\label{Eq_flow_rsv}
   \frac{dR_{v,s}(l)}{dl}=[\eta(l),R_{v,s}(l)],
\end{equation}
with the initial conditions,
\begin{equation}
  R_v(0)=\delta(\bm{r}-\bm{r}'),~~R_s(0)=\beta\delta(\bm{r}-\bm{r}').
\end{equation}

Similar to $H(l)$ in Eq.~\eqref{Eq_H_division}, the $R_{v,s}(l)$ can be also split up into the even $R_{v,s}^e(l)$ and the odd $R_{v,s}^o(l)$ parts,
\begin{equation}
  R_{v,s}(l)=R_{v,s}^e(l)+R_{v,s}^o(l).
\end{equation}
By solving the flow equations \eqref{Eq_flow_rsv} with the perturbative expansion in $1/\tilde{M}$,
we obtain
\begin{subequations}\label{Eq_Rvs}
  \begin{align}
    R_v^e(\infty)=&\delta(\bm{r}-\bm{r}')-\frac{[\mathcal{O}(0),[\mathcal{O}(0),\delta(\bm{r}-\bm{r}')]]}{8\tilde{M}^2} + O(\tilde M^{-3}),\label{Eq_Rvsa}\\
    R_s^e(\infty)=&\beta R_v^e(\infty)-\beta\delta(\bm{r}-\bm{r}')\mathcal{O}(0)\frac{1}{2\tilde{M}^2}\mathcal{O}(0) \nonumber \\
    &-\beta\frac{[\mathcal{O}(0),\delta(\bm{r}-\bm{r}')]}{2\tilde{M}^2}\mathcal{O}(0) + O(\tilde M^{-3}). \label{Eq_Rvsb}
  \end{align}
\end{subequations}
As a result, the vector and scalar densities read,
\begin{equation}
  \rho_{v,s}=\lim_{l\rightarrow\infty}\int d^3r'~\phi^\dag(\bm{r}')R_{v,s}^e(l)\phi(\bm{r}').
\end{equation}

For spherical nuclei, the initial condition for $\mathcal{O}(0)$ [see Eq.~\eqref{Eq_inital_O}] is reduced as
\begin{equation}
   \mathcal{O}(0)=
   \begin{pmatrix}
      0  &-\frac{d}{d r}+\frac{\kappa}{r}\\
      \frac{d}{d r}+\frac{\kappa}{r} &0
   \end{pmatrix},
\end{equation}
where $\kappa=(-1)^{j+l+1/2}(j+1/2)$ depending on the calculated single-particle state.
Accordingly, the reduced nonrelativistic Hamiltonian for nucleons in the Fermi sea reads
\begin{equation}\label{Eq_HF}
  \begin{split}
    H^F=\,&V+S-\frac{d}{d r}\frac{1}{2\tilde{M}}\frac{d}{d r}+\frac{1}{2\tilde{M}}\frac{\kappa(\kappa+1)}{r^2}-\frac{V'-S'}{4{\tilde{M}}^2}\frac{\kappa}{r}+\frac{V''+S''}{8{\tilde{M}}^2}\\
    &-p^2\frac{1}{8{\tilde{M}}^3}p^2-\frac{3S'^2}{16{\tilde{M}}^3}-\frac{S'V'}{8{\tilde{M}}^3}+\frac{V'^2}{16{\tilde{M}}^3} + O(\tilde M^{-4}),
  \end{split}
\end{equation}
with the definition $p^2=-\frac{d^2}{d r^2}+\frac{\kappa(\kappa+1)}{r^2}$.
Furthermore, the vector $\rho_v$ and scalar $\rho_s$ densities can be calculated via
\begin{subequations}
  \begin{align}
     4\pi r^2\rho_v(r)= \,&\rho_0 + \frac{d}{dr}\left[\frac{1}{4\tilde{M}^2}\frac{\kappa}{r}\rho_0\right]+\frac{d^2}{dr^2}\left[\frac{1}{8\tilde{M}^2}\rho_0\right] + O(\tilde{M}^{-3}),\label{Eq_rhov_sph}\\
     4\pi r^2\rho_s(r)=\,&4\pi r^2 \rho_v + \frac{1}{2\tilde{M}^2}\phi^\dag\phi'' - \frac{S'}{\tilde{M}^3}\phi^\dag\phi' - \left[\frac{S'}{\tilde{M}^3}\frac{\kappa}{r} + \frac{1}{2\tilde{M}^2}\frac{\kappa(\kappa+1)}{r^2}\right]\rho_0\nonumber\\
     &- \frac{d}{dr}\left[\frac{1}{2\tilde{M}^2}\frac{\kappa}{r}\rho_0 + \frac{1}{2\tilde{M}^2}\phi^\dag\phi'\right] + O(\tilde{M}^{-3}) \label{Eq_rhos_sph},
  \end{align}
\end{subequations}
where $\phi(r)$ is the eigenfunction of $H^F$ in Eq.~\eqref{Eq_HF} and $\rho_0\equiv\phi^\dag\phi$.
The primes and the double primes denote the first- and second-order derivatives with respect to $r$, respectively.

In the present work, the newly proposed nonrelativistic reduction framework is applied to solve the CDFT  self-consistently.
The point-coupling density functional PC-PK1~\cite{ZhaoPC-PK1} is adopted.
The reduced Hamiltonian $H_F$ for each $\kappa$ is solved in a large set of spherical harmonic oscillator basis with the radial quantum number $n_r < 20$.
The obtained results are compared with the ``exact'' solutions of the CDFT, which are obtained by the shooting method~\cite{meng1998NPA} with the box size $R=20$ fm and the mesh size $0.1$ fm.
The center-of-mass correction energy is taken into account by the phenomenological formula $E_{\rm c.m.}=-\frac{3}{4}41A^{-1/3}~{\rm MeV}$.

%================Results and discussion====================%
%\section{Results and discussion}

\begin{figure}[!htbp]
  \centering
  \includegraphics[width=0.45\textwidth]{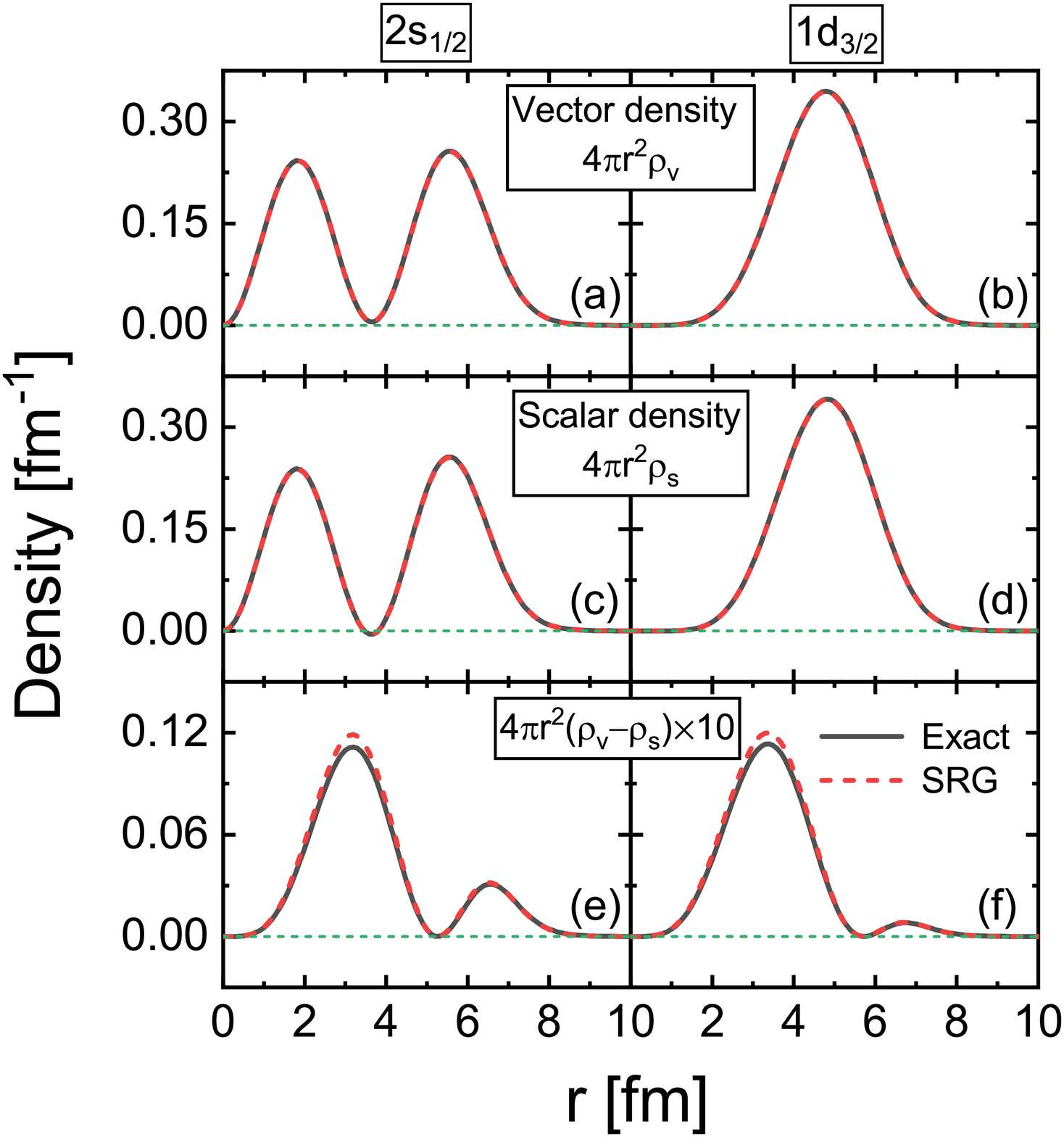}\\
  \caption{(Color online) The vector $4\pi r^2\rho_v$ [panels (a) and (b)] and scalar $4\pi r^2\rho_s$ densities [panels (c) and (d)], as well as their differences $4\pi r^2(\rho_v-\rho_s)$ [panels (e) and (f)] as a function of the radial coordinate $r$ for the single-particle states $2s_{1/2}$ and $1d_{3/2}$ in a given spherical Woods-Saxon potential.
  The solid and dashed lines represent the results given by the exact solutions and the SRG method.
  A factor of 10 has been multiplied for $4\pi r^2(\rho_v-\rho_s)$ in panels (e) and (f).
  }\label{fig1}
\end{figure}

Before performing self-consistent calculations, we first examine the accuracy of the vector and scalar densities in Eqs.~\eqref{Eq_rhov_sph} and \eqref{Eq_rhos_sph} for a fixed spherical Woods-Saxon potential, which corresponds to the neutron potential of $^{208}$Pb in Ref.~\cite{Ren2019SRG}.
In Fig.~\ref{fig1}, the vector $4\pi r^2\rho_v$ and scalar $4\pi r^2\rho_s$ densities, as well as their differences $4\pi r^2(\rho_v-\rho_s)$ for the single-particle states $2s_{1/2}$ and $1d_{3/2}$ are shown as a function of the radial coordinate $r$ in comparison with the corresponding exact solutions.
One can see that both the scalar and vector densities given by the SRG method are almost identical with the exact ones.
The single-particle states $2s_{1/2}$ and $1d_{3/2}$ form a pair of pseudospin doublets, so the small components of the corresponding Dirac spinors share similar distributions and the same number of nodes~\cite{ginocchio1997pseudospin}.
This is revealed by the differences between the vector and scalar densities shown in Figs.~\ref{fig1}(e) and \ref{fig1}(f).
In comparison with the exact solutions, the position of the node is reproduced quite well, and only very slight deviations on the maximum amplitude are found.
This indicates that the present expansion for solving the flow equations on the vector- and scalar-density operators converges accurately and efficiently.
It is worthwhile to mention that such relativistic corrections to the densities have been discussed with the Douglas-Kroll-Hess transformation~\cite{Douglas1974DKH, Hess1986DKH} in Coulombic systems without scalar potentials, and the reconstituted Foldy-Wouthuysen transformation with large scalar potentials~\cite{Guo2019FW, Guo2020FW_density}, where a similar nonrelativistic expansion of $1/\tilde{M}$ is adopted.

\begin{table}
  \centering
  \caption{The total energy $E_{\rm tot}$, the traces of vector Tr$(\rho_v)$ and scalar Tr$(\rho_s)$ densities, and the root-mean-square radius $R_m$ for the nucleus $^{208}$Pb obtained by the exact solutions, the SRG method, and the SRG method with the higher-order ($p^6$) term (see text).
  }\label{tab1}
  \begin{tabular}{ccccc}
    \hline
    \hline
                              &$E_{\rm tot}$ [MeV]    &Tr$(\rho_v)$        &Tr$(\rho_s)$   &$R_m$ [fm]\\
    \hline
       Exact                  &$-1637.45$             &$208.00$            &$199.63$       &$5.62$\\
       SRG                    &$-1652.87$             &$208.00$            &$199.50$       &$5.70$\\
       SRG ($p^6$)            &$-1640.77$             &$208.00$            &$199.60$       &$5.71$\\
    \hline
    \hline
  \end{tabular}
\end{table}

Self-consistent CDFT calculations have been performed with the SRG method by taking the nucleus $^{208}$Pb as an example.
The calculated bulk properties including the total energy, the traces of vector and scalar densities, and the root-mean-square radius are summarized in Table~\ref{tab1} in comparison with the exact solutions, where the traces of vector and scalar densities are calculated by
\begin{equation}
 {\rm Tr}(\rho_{v,s})=\int d^3r~\rho_{v,s}(\bm{r})=\int_0^{\infty}dr~4\pi r^2 \rho_{v,s}(r).
\end{equation} 
All these bulk properties are reproduced satisfactorily with the SRG method.
In particular, the traces of the vector and scalar densities are conserved quite well.

The accuracy of the present SRG method can be improved straightforwardly by taking into account the higher-order terms for $H^F$.
In particular, as mentioned in Ref.~\cite{Ren2019SRG}, the first-order relativistic correction of the kinetic energy, given by $-\beta\mathcal{O}^2(0)\frac{1}{8\tilde{M}^3}\mathcal{O}^2(0)$, can improve the accuracy of the description significantly.
Inspired by the nonrelativistic expansion of the relativistic kinetic energy,
\begin{equation}
   \sqrt{p^2+M^2}-M=\frac{p^2}{2M}-\frac{p^4}{8M^3}+\frac{p^6}{16M^5}+\cdots,
\end{equation}
here, we further take into account the second-order correction for the kinetic energy, i.e.,
\begin{equation}\label{Eq_p6_term}
  \beta\mathcal{O}^3(0)\frac{1}{16\tilde{M}^5}\mathcal{O}^3(0),
\end{equation}
which is denoted as $p^6$-\textit{term} in short.
It is found that the calculated total energy for $^{208}$Pb is improved remarkably by considering the $p^6$-\textit{term} in the present SRG scheme.
The radius is not sensitive to this kinetic energy term, and its improvement should require higher-order terms of $R_{v,s}^e(\infty)$ in Eqs.~\eqref{Eq_Rvsa} and \eqref{Eq_Rvsb}.

\begin{figure}[!htbp]
  \centering
  \includegraphics[width=0.45\textwidth]{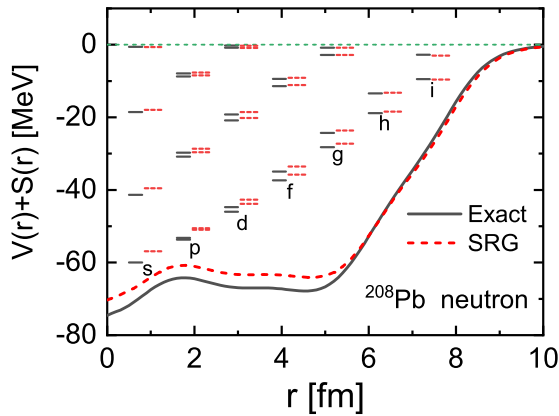}\\
  \caption{(Color online) The self-consistent Kohn-Sham potential $V(r)+S(r)$ and the single-particle spectrum for neutrons in $^{208}$Pb.
  The solid and dashed lines denote the results given by the exact solutions and the SRG method, respectively.
  }\label{fig2}
\end{figure}

Apart from the bulk properties, it is also interesting to compare other quantities.
Figure~\ref{fig2} depicts the self-consistent Kohn-Sham potential $V(r)+S(r)$ and the single-particle spectrum for neutrons in $^{208}$Pb given by the exact solutions and the SRG method.
The overall shape of the Kohn-Sham potential $V(r)+S(r)$ given by the SRG method is in a good agreement with the exact solution.
Quantitatively, the potential obtained by the SRG method is shallower by around $3$ MeV for $r<5$~fm and slightly deeper for $r>6$~fm, in comparison with the exact potential.
It influences the single-particle spectrum accordingly and,  as seen in Fig.~\ref{fig2}, the deeply bound single-neutron levels given by the SRG method are higher than the exact ones by around 3 MeV, while the weakly bound levels are in good agreement with the exact ones.
This is due to the fact that the weakly bound levels are usually more diffuse and, thus, less sensitive to the interior part of the potential, as compared with the deeply bound levels.
In despite of the visible deviations from the exact single-particle levels, it should be noted that the spacings of single-particle energies, especially the spin-orbit splitting energies are reproduced quite well with the SRG method.
This indicates that the nuclear shell structure in the CDFT is preserved quite well by the present nonrelativistic reduction with the SRG method.

It has been found in the solutions with a fixed potential that the exact energies for deeply bound states are reproduced better than those for the weakly bound states, because the latter are usually sensitive to higher momentum components~\cite{Ren2019SRG}. Nevertheless, here the self-consistent single-particle energies exhibit a different feature because they are mainly determined by the interplay between the attractive scalar and repulsive vector potentials. In the present work, we find that the differences between the vector and scalar densities from the SRG approach are slightly larger than the exact solutions in the interior region [see Figs.~\ref{fig1}(e) and \ref{fig1}(f) for the $2s_{1/2}$ and $1d_{3/2}$ densities as examples]. Therefore, as shown in Fig.~\ref{fig2}, the mean potential from the present SRG method is slightly shallower than the exact one in the interior region.

\begin{figure}[!htbp]
  \centering
  \includegraphics[width=0.45\textwidth]{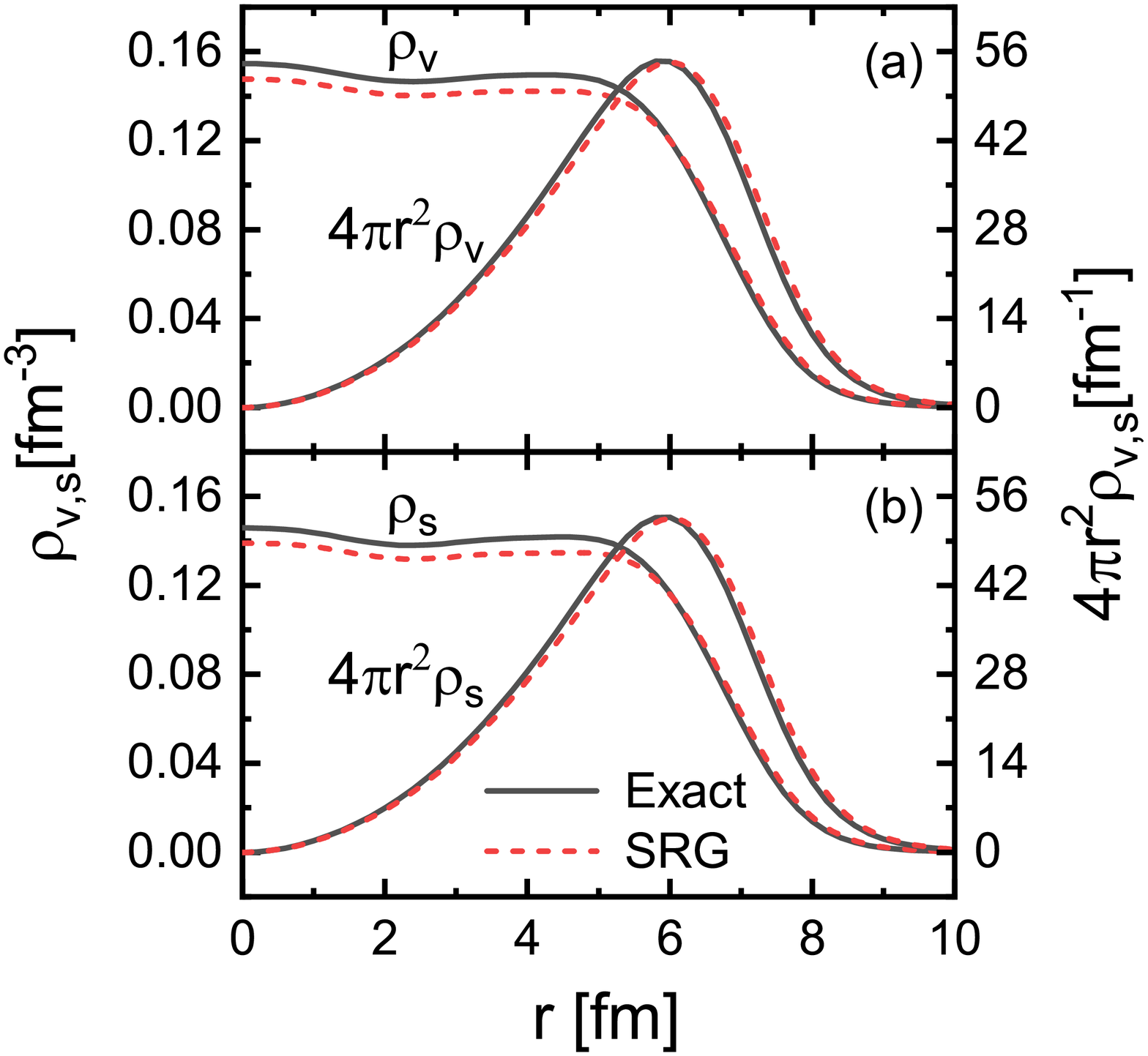}\\
  \caption{(Color online) The total vector $\rho_v$ (top) and scalar $\rho_s$ (bottom) densities for $^{208}$Pb given by the exact solutions (solid lines) and the SRG method (dashed lines).  The densities taking into account the spherical metric factor $4\pi r^2$ are also shown for comparison.
  }\label{fig3}
\end{figure}

In Fig.~\ref{fig3}, the total vector $\rho_v$ and scalar $\rho_s$ density distributions for $^{208}$Pb are presented.
Similar to the Kohn-Sham potentials shown in Fig.~\ref{fig2}, a good overall agreement between the exact densities and the ones given by the SRG method is achieved.
The magnitudes of both scalar and vector densities are underestimated by about 0.007~fm$^{-3}$ in the interior region ($r<5$~fm), while slightly overestimated in the surface part ($r>6$~fm).
This feature is also consistent with that observed in the comparison of the potentials $V(r)+S(r)$.
To calculate the observables, however, one needs to integrate the densities over $r$ from zero to infinity with the spherical metric factor $4\pi r^2$.
One can easily seen that the differences of densities given by the SRG method and the exact solutions are suppressed significantly by the spherical metric.
This is also consistent with the high accuracy achieved in the description of the observables including the total energy and radius listed in Table~\ref{tab1}.

\begin{figure}[h]
  \centering
  \includegraphics[width=0.45\textwidth]{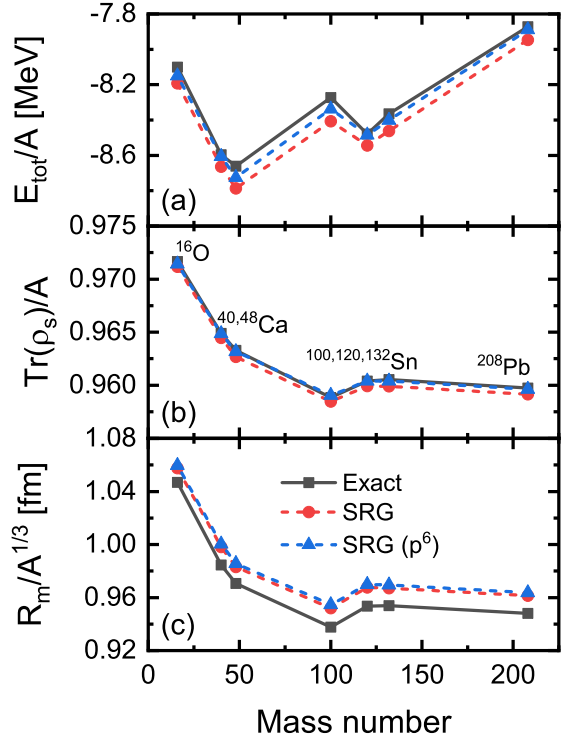}\\
  \caption{(Color online) The total energy per particle $E_{\rm tot}/A$ (top), the trace of scalar density per particle Tr$(\rho_s)/A$ (middle),
  and the root-mean-square radius $R_m$ scaled by $A^{1/3}$ (bottom) for the nuclei $^{16}$O, $^{40,48}$Ca, $^{100,120,132}$Sn, and $^{208}$Pb.
  The squares, circles, and triangles represent the results given by the exact solutions, the SRG method, and the SRG method with the higher-order ($p^6$) term, respectively.
  }\label{fig4}
\end{figure}

Apart from the nucleus $^{208}$Pb, the present SRG scheme has also been examined with other spherical nuclei including $^{16}$O, $^{40,48}$Ca, and $^{100,120,132}$Sn.
The calculated results can be seen in Fig.~\ref{fig4}, where the total energies per particle $E_{\rm tot}/A$, the traces of scalar density per particle Tr$(\rho_s)/A$, and the root-mean-square radii $R_m$ scaled by $A^{1/3}$ are shown in comparison with the exact solutions.
The exact results for the all these spherical nuclei can be reproduced with a satisfactory accuracy.
This indicates that the proposed SRG scheme for the nonrelativistic reduction of the CDFT is applicable for all nuclei.
After taking into account the $p^6$-term in the SRG method, the description of the total energies and the traces of scalar density given by the SRG method is improved for all  nuclei.
The mean deviation for $E_{\rm tot}/A$ is reduced to $0.035$ MeV for the seven nuclei.
Similar to the case of $^{208}$Pb, the radii are not sensitive to the $p^6$-term for all nuclei, since their improvement should require higher-order terms of $R_{v,s}^e(\infty)$ in Eqs.~\eqref{Eq_Rvsa} and \eqref{Eq_Rvsb}.

%===============SUMMARY ==============

%\section{Summary}
In summary, the nonrelativistic reduction of the self-consistent covariant density functional theory is realized for the first time with the similarity renormalization group method.
The reduced nonrelativistic Hamiltonian is obtained by solving the Dirac Hamiltonian flow equations with a novel expansion in terms of the inverse of the Dirac effective mass.
The vector and scalar densities are determined by the corresponding transformation of the density operators.
The efficiency and accuracy of this newly proposed framework have been demonstrated for several typical spherical nuclei.
It is found that the exact solutions of the total energies, traces of vector and scalar densities, and the root-mean-square radii are reproduced quite well for all nuclei.
By taking $^{208}$Pb as an example, the Kohn-Sham potential, the single-particle spectrum, and the density distributions are examined, and a satisfactory agreement with the corresponding exact solutions is achieved.
This allows one to directly compare and bridge the relativistic and nonrelativistic nuclear energy density functional theories in the future.

%\begin{acknowledgments}
This work was partly supported by the National Key R\&D Program of China (Contracts No. 2018YFA0404400 and No. 2017YFE0116700), the National Natural Science Foundation of China (Grants No. 11621131001, No. 11875075, No. 11935003, and No. 11975031), and the State Key Laboratory of Nuclear Physics and Technology, Peking University (No. NPT2020ZZ01).
%\end{acknowledgments}

%\bibliography{reference}

\end{document}